\documentclass[sigconf]{acmart}

\copyrightyear{2022}
\acmYear{2022}
\setcopyright{rightsretained}
\acmConference[C\&C '22]{Creativity and Cognition}{June 20--23, 2022}{Venice, Italy}
\acmBooktitle{Creativity and Cognition (C\&C '22), June 20--23, 2022, Venice, Italy}\acmDOI{10.1145/3527927.3533735}
\acmISBN{978-1-4503-9327-0/22/06}




\begin{document}

\title{Redefining Access to Large Audiovisual Archives through Embodied Experiences in Immersive Environments}
\subtitle{Creativity \& Cognition 2022 - Graduate Student Symposium}
\author{Giacomo Alliata}
\email{giacomo.alliata@epfl.ch}
\date{April 2022}

\begin{abstract}
    Audiovisual archives are the mnemonic archives of the 21st century, with important cultural institutions increasingly digitizing their video collections. However, these remain mostly inaccessible, due to the sheer amount of content combined with the lack of innovative forms of engagement through compelling frameworks for their exploration. The present research therefore aims at redefining access to large video collections through embodied experiences in immersive environments. The author claims that, once users are empowered to be actors of the experience rather than mere spectators, their creativity is stimulated and narrative can emerge.
\end{abstract}

\maketitle

\section{Introduction}
Audiovisual archives are the visual memories of the 21st century, with large cultural collections being entirely digitized, such as the Radio Télévision Suisse with their 200000 hours of footage or the BCC with more than a million hours. Cultural video collections are furthermore useful to preserve intangible cultural heritage (ICH), "the culture that people practise as part of their daily lives" (\cite{kurin}). Lenzerini talks of "all [the] immaterial manifestations of culture [that] represent the variety of living heritage of humanity as well as the most important vehicle of cultural diversity" (\cite{lenzerini}). ICH being difficult to safeguard, audiovisual archives are commonly used to keep a trace of rituals, sports and musical performances, to name a few meaningful examples.

However, these large archives remain mostly inaccessible, due to the sheer amount of content combined with the lack of innovative forms of engagement through compelling frameworks for their exploration. The importance of the latter has also been stressed by archival scholars (\cite{fossati-2012}), revealing a clear gap in the field for the specific case of audiovisual materials. The Digital Humanities community and the GLAM sector are in need of innovative ways to approach video collections, so that they can be presented to the general public.

\section{Research}

\subsection{Aim}
Within the larger scope of the interdisciplinary project \emph{Narratives from the Long Tail: Transforming Access to Audiovisual Archives}\footnote{SNSF Sinergia project, grant number CRSII5\_198632, see \url{https://www.epfl.ch/labs/emplus/projects/narratives/} for a project description.}, this research will investigate how large video collections can be explored in a compelling way for the general public. Both a general framework and concrete demonstrators applied to specific archives are expected to be produced.

\subsection{Theoretical background}
The present research is at the intersection of human-computer interaction, immersive environments, the functionality of the application and embodiment theories. 
The concepts of embodiment and its dimensions (\cite{johnson-2008}), as well as founding theories of immersion and interactivity (\cite{aylett-1999, milgram-1994}) and their more recent revisions (\cite{skarbez-2021, rubio-2017}) are elaborated to understand how narrative can emerge in such frameworks (\cite{kenderdine-2015}). The duality of immersion and presence in immersive environments is studied in relationship to various forms of narrative (\cite{brown-2011}). The present research also leverages how data can be used as material for the creation of a virtual world in which narrative can emerge through users' interactions (\cite{kenderdine-2013}). Furthermore, the importance of the social relations that arise in multi-users environments, placing users as actors of the storytelling rather than mere spectators, will be discussed, highlighting the clear benefits in terms of enjoyment of the experience and understanding of the cultural aspect, as well as the key creative aspect of such interactive installations.

\subsection{Expected contribution}
The main contribution will be a novel framework to explore large audiovisual archives through the emergence of narrative in immersive environments. Alongside this general framework, concrete demonstrators will be developed to explore four collections: the Radio Télévision Suisse, the International Olympics Committee, the Montreux Jazz Digital Project and the Netherlands Eye Filmmuseum Mutoscope and Biograph Collection. These four audiovisual archives represent hundreds of thousands of hours of footage and are currently lacking compelling systems to discover them, especially for the general public. Only a small portion of the RTS archive is for instance available online, and only through a classic database navigable with textual and categorical search tools\footnote{RTS archive: \url{https://www.rts.ch/archives/}}.

\begin{figure}[!h]
    \centering
    \includegraphics[width = 0.3\textwidth]{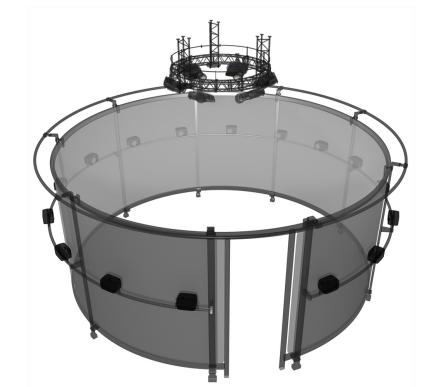}
    \caption{Schematic view of the Panorama+ omnidirectional stereographic immersive environment (credits: Sarah Kenderdine)}
    \Description{Schematic view of the Panorama+ omnidirectional stereographic immersive environment}
    \label{fig:panorama}
\end{figure}

These systems will mainly be built for the Panorama+, an omnidirectional stereographic immersive environment (\ref{fig:panorama}), shown to enhance cognitive exploration and interrogation of high-dimensional data by placing users inside a dataset (\cite{shen-2019}), but the use of other immersive technologies will also be explored. This visualization system is based on the UNSW iCinema Research Centre’s landmark 360-degree stereoscopic interactive visualisation environment - AVIE (Advanced Visualisation and Interaction Environment, \cite{avie}).

\subsection{Relevance to Creativity \& Cognition 2022}

This research puts forward the idea that creativity can arise in these types of media installations, in which users are invited to embrace an acting role rather than being passive spectators. By driving the behaviour of the interactive installation, visitors become agents and create their own narrative, for themselves and for the rest of the audience. This opposes the traditional views that assume creativity cannot be learned or trained, claimed to only be an innate ability some individuals have and others do not (\cite{murray, efland, goldstein}). However, as Wilf demonstrates, "ethnographic records suggests that creative agency results from socially informed and consequential processes of socialization" (\cite{eitan_wilf}). In particular, (a) imitation socializes participants into creativity (\cite{ingold_hallam}), (b) external inputs can shape ones' spontaneous interiority (\cite{mahmood, wilf_2012}) and (c) first-time users first need to learn the boundaries of the emergent frames they are expressing themselves in before being able to "breach expectations" and generate new means of creativity (\cite{howard}). In the frame of multi-users interactive installations, visitors receive inputs from both the system and the other persons in the space, consciously or not, and perform interactions that generate a response, therefore creating a new narrative, within the boundaries set by the functionalities of the application. Furthermore, the latest theoretical frameworks on creativity stress the importance of public (\cite{sternberg-2022}) because upon the judgement of others depend the novelty and value of the particular arrangement generated, the founding elements of creativity itself (\cite{sternberg-2022-missinglinks, sternberg-2022}). In the case of installations depending on users' interactions (to trigger some kind of generative behavior for instance), Geert Mul's "third person's perspective" thus becomes essential (\cite{mul-2018}).

\begin{figure}[!h]
    \centering
    \includegraphics[width = 0.4\textwidth]{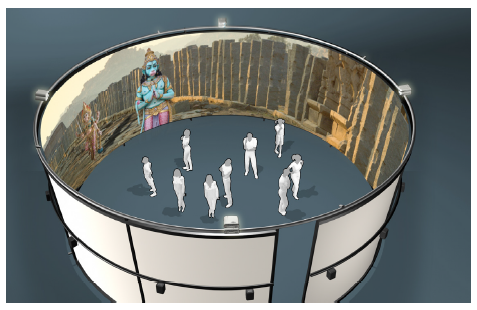}
    \caption{Schematic view of the installation \emph{PLACE-Hampi}  (credits: Sarah Kenderdine)}
    \Description{Schematic view of \emph{PLACE-Hampi}}
    \label{fig:place-hampi}
\end{figure}

Creativity has also been shown to have a key role in assimilating knowledge, as put forward by constructionist theories (\cite{dewey-1966, piaget-1973}), according to which individuals do not learn by passively perceiving content but rather by actively crafting, manipulating and therefore creating new knowledge. Through the concept of embodiment in immersive environments, this is made possible even in virtual worlds, where various approaches can be taken to generate a sort of dialogue with the users. An interesting example can be found in the interactive installation \emph{PLACE-Hampi} (\ref{fig:place-hampi}), an "embodied theatre of participation" that "permits an unprecedented level of viewer co-presence in a narrative-discovery of a cultural landscape", facilitating "dynamic inter-actor participation and cultural learning" (\cite{kenderdine-2007}). In this installation that virtually takes places at the World Heritage site Vijayanagara (Hampi), South India, machine agents representing various deities react to users' actions in the space, thus contributing to the emergence of a dialogue between systems and users, driven by the latter. Simultaneously, this references the concept of "darshan", of seeing and being seen by a deity, and grounds even more the experience in its particular cultural setting. Relativistic perspectives stating the importance of the sociocultural context in which creativity is considered (\cite{sternberg-2022}) help to further analyze this work. It is rather obvious that, for most people in Western world, concepts like "darshan" and relationships to Hindu deities will not bear the same weight than for people experiencing this culture in their everyday life, thus altering the way different groups of people will experience this installation.

\section{Completed and ongoing work}

\subsection{Literature review}

This research being at the intersection of multiple areas, only a short summary is provided here, highlighting key theories. 

Although curatorial practices are not a central focus of this work, they are nonetheless an important body of research to consider, considering archival scholars are amongst the ones challenging current access practices to said archives. "Generous interfaces" are required to reveal the true extent and intricacy of large cultural collections (\cite{whitelaw-2015}), since traditional queries based on metadata restrain users to prior interpretations and prevent them to rely on more visual features that are harder to verbally capture (\cite{masson-2020}). Curatorial practices might also be necessary to find the proper cultural collection to work with, and to prune it to the specific needs of the project. In this case, traditional metadata can sometimes be used to select objects, to filter down collections, simultaneously repurposing them from their original goal (\cite{mul-2018}).

Regarding immersion and presence theories, as stated previously, the Taxonomy of Mixed Reality Visual Displays of Milgram and Kishino is a key founding theory to understand the properties and effects on users of various kinds of systems. The three axis used to describe the different displays are the "Extent of World Knowledge", the "Reproduction Fidelity" and the "Extent of Presence Metaphor". Furthermore, a distinction between immersion and presence has to be made (\cite{slater-1997}): the first is a direct consequence of the technology employed, while the second can be regarded as a state of consciousness, that is not directly dependent on the system (a more immersive system does not imply that users will feel more present, as other factors have to be taken into account). More recent taxonomies build on top of Slater's ideas and highlight key dimensions such as embodiment or interactivity (\cite{ruscella-2021}).

To elaborate the idea of embodiment, Johnson defines five dimensions of the body: the body as a biological organism, the ecological, the phenomenological, the social and the cultural body (\cite{johnson-2008}). Although these are defined as psychological concepts, without any reference to immersive environments, the question of embodiment is key in this kind of visual displays, as the degree of embodiment in an experience can affect the users' sense of presence. This claim has vastly been explored by Jeffrey Shaw and Sarah Kenderdine for instance, with works like \emph{PLACE-Hampi}, previously described, \emph{Pure Land AR}\footnote{\emph{Pure Land AR}: \url{https://sarahkenderdine.info/installations-and-curated-exhibitions/pure-land-ar}} or \emph{Conversations}\footnote{\emph{Conversations}: \url{https://www.jeffreyshawcompendium.com/portfolio/conversations/}}.

Within the visualisation community, the link between users creating new knowledge and the modes of interaction they have access to is a core idea. There is indeed a direct dependence between users' objectives and the affordances the system offers (\cite{pike-2009}). Moreover, the functionalities of interaction present in an installation directly influence how users can create new narratives (\cite{aylett-1999}). It is clear that in the extreme case of a simple movie for instance, where users are only passive spectators without any means of controlling the unfolding of events, the only narrative present is the one predefined by the movie creator. There is no creative aspect, at least not immediate and with an effect on the movie itself. On the other hand, a system that reacts to users' inputs is able to generate new narratives, up to a certain degree, and therefore users become active creators of the experience. Particularly interesting is the dual view offered by Geert Mul. In his work, users can both be seen as "highly productive, in that the appearance of the works changes based on their input" and "merely one in a much larger series of variables that determine the outcome of the calculation" (\cite{mul-2018}). Although these two views might seem at odds and attribute a different weight on users' role, their impact is nonetheless real and creative, and their interpretative role of the system output is always crucial. It is worth mentioning that this also has a direct consequence on the social interactions between people sharing the same experience in a multi-users environments. In this case, various modes of relationships can manifest: a single user might be driving the experience for a larger audience (that will experience Mul's aforementioned "third person's perspective"), the actions of each user could have an impact on the virtual world and thus be perceived by the other persons directly in the virtual world, or people could even communicate and interact within the virtual world, through the use of avatars. 

Finally, all these concepts are necessary to understand how different kinds of interactive narratives can emerge in immersive environments, and how users can drive and create their own stories while exploring large audiovisual archives. An interesting case is the so called "transcriptive narrative", where data is used as a material that users are able to sculpt, modify and recombine to create new meanings (\cite{brown-2011}). As a concrete example, one can consider the installation \emph{T\_Visionarium II}\footnote{\emph{T\_Visionarium II}: \url{https://www.jeffreyshawcompendium.com/portfolio/t_visionarium-ii/}} in which television footage is segmented into short clips and manually tagged, so that users are able to recombine them at will and thus create new stories.

\subsection{Claude Nicollier Video Archive}

\begin{figure}[!h]
    \centering
    \includegraphics[width=0.45\textwidth]{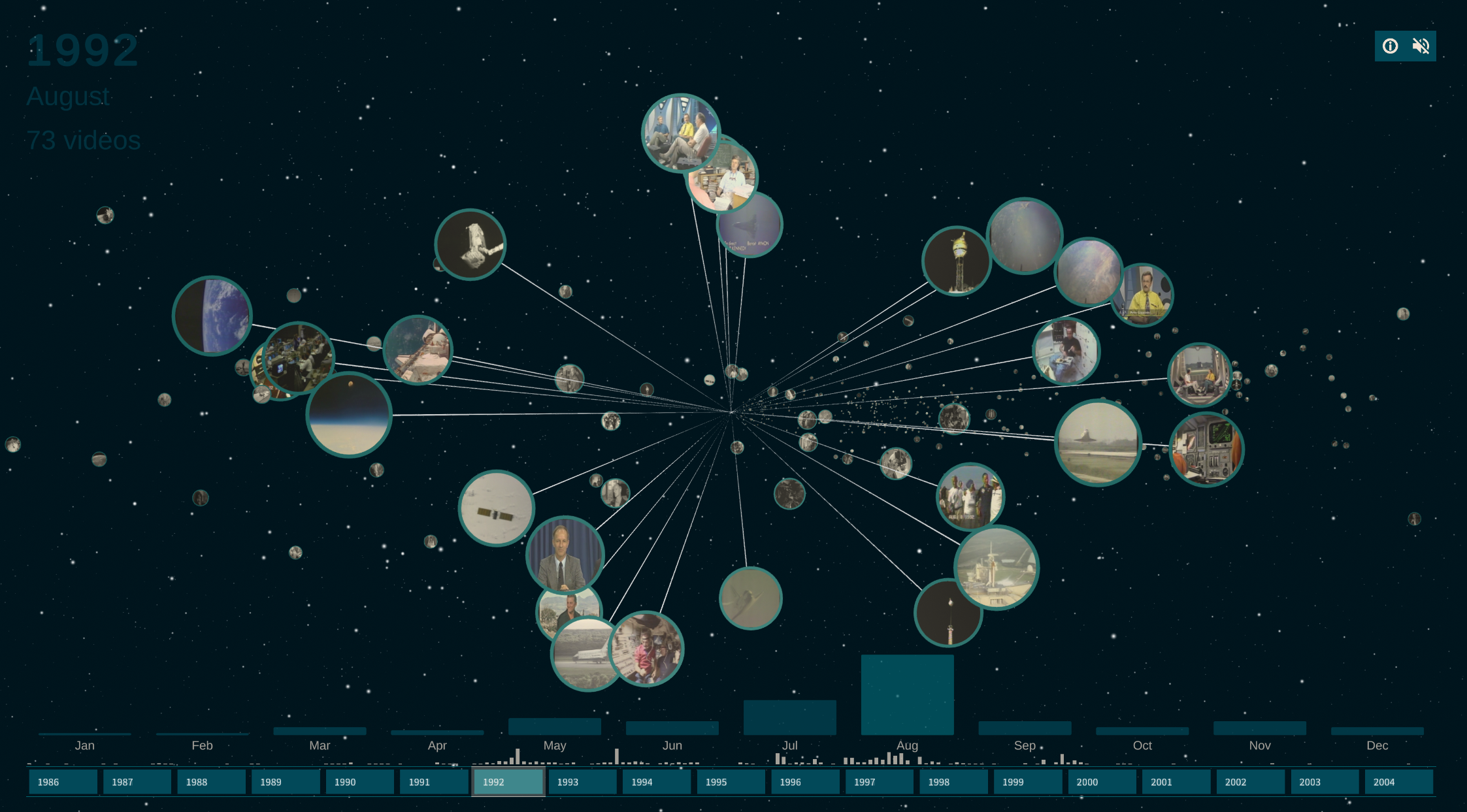}
    \caption{Main view of the installation \emph{Orbiting Memories}, with orbiting videos of the current year on the center and navigation timeline on the bottom}
    \Description{Main view of the application, with orbiting videos of the current year on the center and navigation timeline on the bottom}
    \label{fig:orbiting_memories_mainview}
\end{figure}

\begin{figure}[!h]
    \centering
    \includegraphics[width=0.45\textwidth]{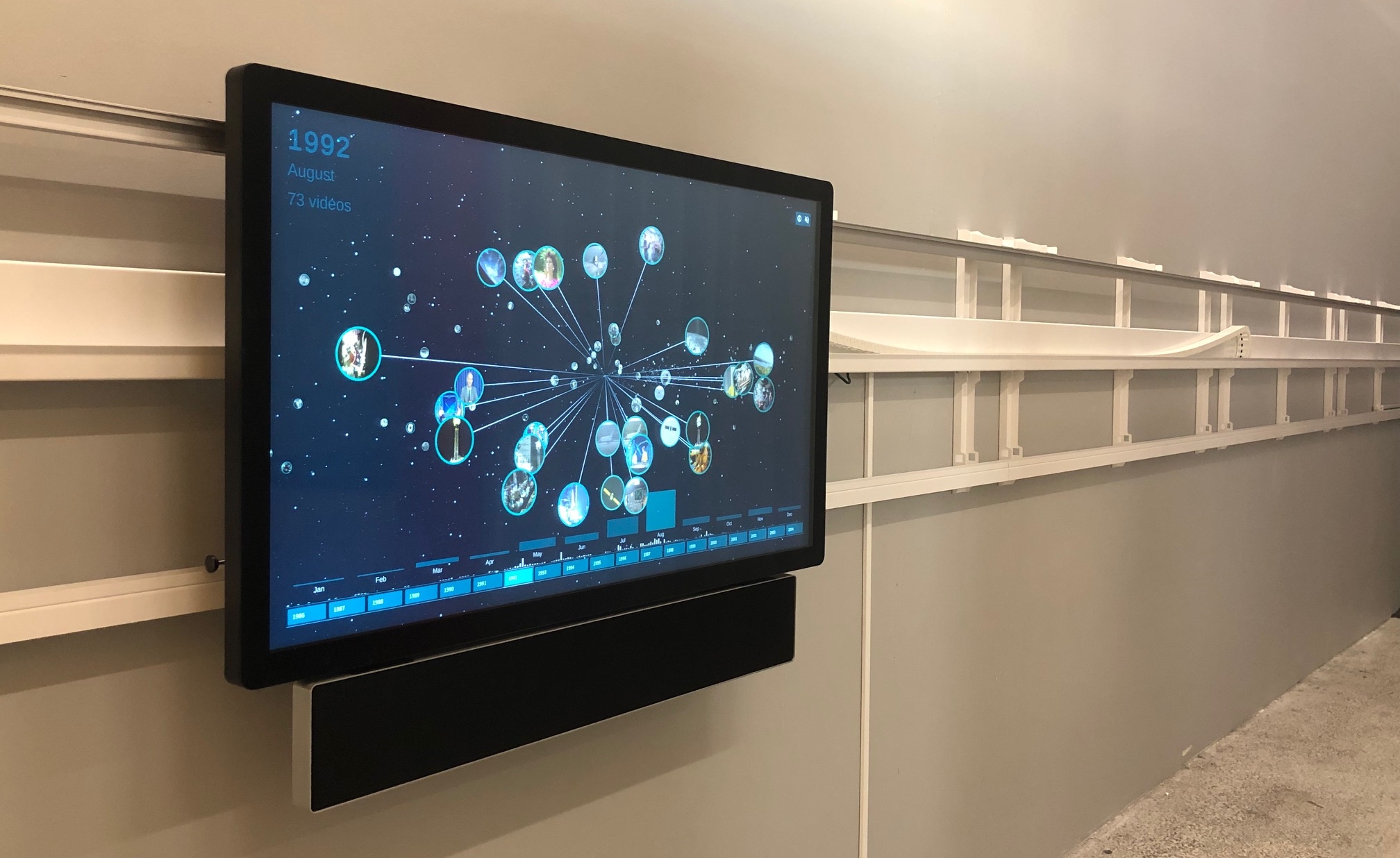}
    \caption{View of the installation \emph{Orbiting Memories} on the Linear Navigator, showing a portion of the rail the screen moves on}
    \Description{View of the Linear Navigator, showing a portion of the rail the screen moves on}
    \label{fig:orbiting_memories_ln}
\end{figure}

Alongside the literature review, a concrete installation has been built, to explore an interesting video collection.
The Claude Nicollier Video Archive\footnote{Claude Nicollier Video Archive: \url{https://www.epfl.ch/innovation/domains/cultural-heritage-and-innovation-center/claude-nicollier-video-archive/}} is a collection of 600 videos related to Swiss astronaut Claude Nicollier, digitized by EPFL Cultural Heritage and Innovation Center. This archive spans over 20 years and records Nicollier's life, as a member of his family, a Swiss Air Force pilot and a NASA/ESA astronaut.
The resulting installation, \emph{Orbiting Memories}, is an interactive application created to explore in a serendipitous way this collection. It is built on the Linear Navigator, a movable 4k touch screen mounted on a 12 meter rail, where the full range of motion is here mapped to the years from 1985 to 2004, where the bulk of videos lie, as shown in figures \ref{fig:orbiting_memories_mainview} and \ref{fig:orbiting_memories_ln}. The installation will be exposed at the \emph{Cosmos Archeology} exhibition, to be held at EPFL Pavilions, Lausanne\footnote{EPFL Pavilions, Lausanne: \url{https://epfl-pavilions.ch/}}, from September 2022.

The application is built with a 1:1 mapping, in a "window-on-the-world" setting (\cite{milgram-1994}). Users are therefore confronted with a virtual world in which the archival material is organized in orbital movements, grouped by year and month. Hence the title of the installation, making explicit the metaphor proposed to explore memories of Claude Nicollier's life and career as if they were celestial bodies orbiting in space. To compliment this idea, the starry background uses real-world positions of stars, as they would be seen from Earth.

Furthermore, the installation proposes a kinaesthetic experience where users are invited to walk along the rail and follow the screen on its journey through the memories, recalling the idea of the "phenomenological [or] tactile-kinaesthetic body" (\cite{johnson-2008, sheets-1999}). A journey that is specifically created on the fly by users themselves, by selecting specific years on the timeline at the bottom of the screen. Therefore contrasting the linear nature of the mechanical system and of history, users create their own narrative, exploring the collection and reliving Nicollier's career on their own terms.

\section{Planned trajectory towards thesis completion}
The following section presents a planned trajectory towards thesis completion.

\textbf{First year}
\begin{itemize}
    \item Literature review on human-computer interaction, theories of immersion and presence, theories of embodiment, emergence of narrative and relationship with creativity.
    \item Review of media installations in immersive environments to explore large archives, with a focus on visual collections (images and videos).
    \item Creation of \emph{Orbiting Memories}, a digital installation to explore the Claude Nicollier Video Archive.
    \item Initial concepts for an installation to explore the Lausanne Prix collection, an audiovisual archive of dance performances.
    \item Initial concepts for the demonstrators to explore the four archives of the \emph{Narratives from the Long Tail: Transforming Access to Audiovisual Archives} project.
\end{itemize}

\textbf{Second year}
\begin{itemize}
    \item Continuing and updating the literature review.
    \item Complete the Lausanne Prix installation.
    \item Produce a theoretical framework to explore large audiovisual archives.
    \item Implement first demonstrators for the four archives, applying concepts of the theoretical framework.
    \item Academic writings to present the theoretical framework and its application.
\end{itemize}

\textbf{Third year}
\begin{itemize}
    \item Perform users testing sessions on the demonstrators to validate theoretical and conceptual framework.
    \item Improve theoretical framework and its generalization.
    \item Apply framework to new audiovisual archives (an interesting option, either in the third or fourth year, would be to work on the RaiTeche archive, through a residency in Rome at the Instituto Svizzero\footnote{Residencies at Instituto Svizzero: \url{https://www.istitutosvizzero.it/residenze/}}.
    \item Academic writings to present results of the users testing sessions, improvements of the framework and concrete demonstrators built.
\end{itemize}

\textbf{Fourth year}
\begin{itemize}
    \item Finish applying framework to new audiovisual archives.
    \item Prepare exhibition at EPFL Pavilions, Lausanne to showcase the demonstrators built.
    \item Writing thesis.
\end{itemize}

\section{Conclusion}
This paper has presented the current status of research as well as a planned trajectory towards thesis completion. The research lies at the intersection of human-computer interaction, immersive environments, theories of embodiment and emergence of narrative, and the main contribution is expected to be a novel framework to explore large audiovisual archives.

The installation \emph{Orbiting Memories} has also been discussed, with the premises of a theoretical framework linking the different key theories referenced.

Finally, relevance to the Creativity \& Cognition conference has been highlighted, based on the claim that creativity can arise in embodied experiences in immersive environments and contribute to the emergence of new narratives, with clear benefits for the users.

\begin{acks}
The author thanks Professor Sarah Kenderdine for the supervision of this thesis as well as colleagues PhD students Yuchen Yang and Yumeng Hou for their support and insights. 

The other members of the Laboratory of Experimental Museology are also worth of mention, for their technical expertise and moral support.

Finally, designer Patrick Donaldson is thanked for his help on the {Orbiting Memories} installation.
\end{acks}

\bibliographystyle{ACM-Reference-Format}
\bibliography{references}

\end{document}